\documentstyle[12pt,epsf]{article}
\begin{document}
\noindent
\begin{center}
{\LARGE {\bf Conformal Invariance, Accelerating Universe and The
Cosmological Constant Problem\\}} \vspace{2cm}
${\bf Yousef~ Bisabr}$ \footnote{e-mail:~y-bisabr@srttu.edu.}\\
\vspace{0.5cm} {\small {Department of Physics, Shahid Rajaee
University, Lavizan,
Tehran 16788, Iran.}}\\
{\small {Institute for Studies in Theoretical Physics and
Mathematics, Tehran, Iran.}}
\end{center}
\vspace{1cm}
\begin{abstract}
We investigate a conformal invariant gravitational model which is
taken to hold at pre-inflationary era.  The conformal invariance
allows to make a dynamical distinction between the two unit
systems (or conformal frames) usually used in cosmology and
elementary particle physics. In this model we argue that when the
universe suffers phase transitions, the resulting mass scales
introduced by particle physics should have variable contributions
to vacuum energy density. These variations are controlled by the
conformal factor that appears as a dynamical field. We then deal
with the cosmological consequences of this model. In particular,
we shall show that there is an inflationary phase at early times.
At late times, on the other hand, it provides a mechanism which
makes a large effective cosmological constant relax to a
sufficiently small value consistent with observations.  Moreover,
we shall show that the conformal factor acts as a quintessence
field that leads the universe to accelerate at late times.

\end{abstract}
\vspace{2cm}
\section{Introduction}
There is a fundamental conflict between observations and
theoretical estimates on the value of the cosmological constant.
In view of the cosmological observations we have an upper limit
equivalent to $\frac{\Lambda}{G} \sim (0.003~ev)^4$.  On the
other hand, the standard model of particle physics implies that
the universe has undergone a series of phase transitions at early
epoch of its evolution contributing to the vacuum energy density
$120$ order of magnitude larger than this observational bound.
Understanding of such a large discrepancy remains as one of the
main problems of theoretical physics.  There have been many
attempts trying to resolve the problem \cite{1}.  Most of them
are based on the belief that $\Lambda$ may not have such an
extremely small value at all the time and there should exist a
dynamical mechanism working during evolution of the universe
which provides a cancellation of the vacuum energy density at
late times \cite{2, 3}.\\
Among different kinds of such models of a decaying cosmological
constant, more promising ones may be those which consist of a
scalar field nonminimally coupled to gravity which itself is
described by the usual Einstein-Hilbert action with a
cosmological term \cite{3}. The scalar field evolves with cosmic
expansion in such a way that its energy density which is unstable
due to the gravitational interaction compensates the vacuum
energy density. Generic feature of these models is that they
results in an asymptotic behavior $\Lambda \sim t^{-2}$ which in
the present epoch roughly gives the observed upper bound.
Nevertheless, as an immediate consequence of a nonminimal
coupling these models entail an effective gravitational coupling
which also behaves asymptotically as $G_{eff} \propto t^{-2}$,
namely that gravity turns off at the cost of having a small
cosmological constant.\\
This dramatic behavior encourages one to think about theories in
which the gravitational coupling itself appears as a dynamical
field, namely the scalar-tensor theories of gravitation.  Along
this line of thought we have concerned here with a particular
form of these theories which is conformally invariant. The
conformal invariance implies that the theory is invariant under
local changes of units of length and time or local unit
transformations \cite{4}.  In such transformations different unit
systems or conformal frames are related via spacetime dependent
conversion (or conformal) factors.  Thus there exists a dynamical
distinction between any two different unit systems in a conformal
invariant model.  The reason for introducing of such a model to
study the cosmological constant problem is the important fact
that the observational estimates and the theoretical predictions
are actually carried out in two different unit systems, the unit
systems usually used in cosmology and elementary particle
physics\footnote{From now on these are referred to as the
cosmological and the quantum unit systems, respectively.}. It is
generally assumed that these two unit systems are related by a
constant conversion factor.  In other terms they are transformed
by a
global unit transformation.\\
In a local unit transformation, on the other hand, changes of
unit systems find a dynamical meaning.  We have already shown
\cite{bs} that this dynamical changes of unit systems can be taken
as a basis for constructing a cancellation mechanism which reduces
a large effective cosmological constant to a sufficiently small
value.  In the present work, we intend to study the effects of
the model introduced in \cite{bs} both on the early and the late
times asymptotic behaviour of the scale factor in the standard
cosmological model.  In particular, we wish to answer the
questions that how this model affect the inflation at early times
and whether it leads to an accelerating
universe at late times.\\
We shall assume that gravity is described by a conformal invariant
gravitational model at early universe, specifically before
breaking the gauge symmetry of fundamental interactions at GUT
energy scale.  When the gauge symmetry is spontaneously broken
the resulting vacuum energy density as a dimensional parameter
may be considered in different conformal frames.  To deal with
the cosmological implications of such a parameter it is essential
to note that the cosmological and the quantum frames are
dynamically distinct and therefore the vacuum energy density, as
a mass scale introduced by particle physics, should take a
variable configuration in the cosmological frame. This variation
is then controlled by the conformal factor that appears as a
dynamical field in our model. We shall show that this dynamical
field plays a key role in a cancellation mechanism that works
essentially due to expansion of the universe.  It also appears as
a quintessence field which causes the universe to
accelerate at late times.\\
We organize this paper as follows: In section 2, we introduce a
conformal invariant gravitational theory consisting of a real
scalar field which is conformally coupled to gravity. It is the
classical analogue of the model which we have previously
investigated in \cite{11}. In section 3, it is argued that
introduction of an effective cosmological constant to the model
requires that one considers a dynamical distinction between the
cosmological and the quantum unit systems. We then study the
consequences of this distinction in two parts. Firstly,
consideration of the model at early times reveals that it can
bring the universe into an inflationary phase.  Secondly, we
shall show that while the model leads to a damping behavior for
the effective value of the cosmological constant at late times,
it avoids the aforementioned problem on the gravitational
coupling.  Moreover, evolution of the scale factor indicates that
the model gives rise to an accelerating expansion for the universe
at late times.  In section 4, we
summarize and discuss our results.\\
Throughout this paper we work in units
in which $\hbar=c=1$ and the sign conventions are those of MTW \cite{a12}.\\
\section{The model}
We shall consider a gravitational system which consists of a scalar field
$\phi$ and the gravitational field, described
by the action\footnote{This action has been investigated in different
contexts.  See, for example, \cite{11, 15}.}
\begin{equation}
S=-\frac{1}{2} \int d^{4}x \sqrt{-g}~ (g^{\mu\nu} \nabla_{\mu}\phi
\nabla_{\nu}\phi+\frac{1}{6} R \phi^{2}) \label{2.1}\end{equation}
where $\nabla_{\mu}$ denotes a covariant differentiation and $R$
is the Ricci scalar.  Note that the action is not involved the
free gravitational field contribution.  The gravitational
coupling is here dynamic and is given by $\sim \phi^{-2}$.\\  The
remarkable feature of (\ref{2.1}) is that it is invariant under
conformal transformations
\begin{equation}
g_{\mu\nu} =e^{2\sigma}  \bar{g}_{\mu\nu}
\label{2.6}\end{equation}
\begin{equation}
\phi(x) = e^{-\sigma}  \bar{\phi}(x) \label{2.7}\end{equation}
where $\sigma$ is a smooth dimensionless spacetime function. This
means that the theory described by the action (\ref{2.1}) can be
described in many different conformal frames which are
dynamically equivalent.  They correspond to various configurations
one assigns to the scalar field $\phi$ or various choices of
local standards of units.  Therefore different conformal frames
may be distinguished by local values of some
dimensional parameters which enter the theory.\\
Variation of (\ref{2.1}) with respect
to $g^{\mu\nu}$ and $\phi$ yields, respectively,
\begin{equation}
G_{\mu\nu}=-6\phi^{-2}(\nabla_{\mu }
\phi \nabla_{\nu}\phi-\frac{1}{2}g_{\mu \nu} \nabla_{\gamma  }\phi
\nabla^{\gamma  }\phi)+\phi^{-2}(\nabla_{\mu }\nabla_{\nu}-g_{\mu\nu}\Box
)\phi^2
\label{2.2}\end{equation}
and
\begin{equation}
\Box \phi -\frac{1}{6}R \phi=0 \label{2.3}\end{equation} Here
$\Box \equiv g^{\mu \nu} \nabla_{\mu } \nabla_{\nu}$ and $G_{\mu
\nu}$ is the Einstein tensor.  One should recognize that the
equations (\ref{2.2}) and (\ref{2.3}) are not independent.
Indeed, the trace of (\ref{2.2}) gives
\begin{equation}
\phi(\Box-\frac{1}{6}R)\phi=0 \label{2.5}\end{equation} which
contains the equation (\ref{2.3}).  This is a direct consequence
of the absence of a dimensional parameter in the model.  In the
next section we shall introduce a cosmological constant which
leads the field equations to be independent.  Note that addition
of a dimensional parameter in a conformal invariant model would
not necessarily mean the breakdown of the symmetry since in a
local unit transformation all the dimensional parameters are
expected to be transformed according to their dimensions. The
conformal invariance is broken when one singles out a particular
conformal frame in which the dimensional parameter or the
gravitational
coupling $\phi^{-2}$ take on preferred constant values.\\
\section{Cosmological implications}
\subsection{Vacuum-dominated era}
It is generally believed that at GUT energy scale the universe
has passed through a certain disordered phase associated with the
gauge symmetry of the grand unified theories.  When the gauge
symmetry is spontaneously broken the structure of the vacuum
drastically changes in the sense that it acquires a large amount
of energy density which appears as a large effective cosmological
constant. The key question which should be answered at this stage
is that how this vacuum energy density should be coupled to gravity.\\
To clarify this point we remark that the use of two different
unit systems are conventional for measuring this energy density.
On one hand, the upper bound set by observations is obtained in a
unit system which is defined in terms of large scale cosmological
parameters (the cosmological unit system).  On the other hand,
the theoretical predictions are based on a natural unit system
which is suggested by quantum physics (the quantum unit system).
One usually presupposes that these two unit systems should be
indistinguishable up to a constant conversion factor in all
spacetime points.  It means that they should transform to each
other by a global unit transformation.  Such a global
transformation clearly carries no dynamical implications and the
use of a particular unit system is actually a matter of
convenience.\\
Here we would like to consider a different approach.  We
introduce a theoretical scheme in which an explicit recognition
is given to the distinguished characteristics of the cosmological
and the quantum unit systems.  In such a theoretical scheme  one
should no longer accept the triviality one usually assigns to a
unit transformation.\\
We first assume that gravity is described by the conformal
invariant gravitational model (\ref{2.1}), before the universe
goes through phase transition. In the context of this conformal
invariant model we intend to consider local unit transformations.
In fact, the conformal invariance of (\ref{2.1}) implies that
there exists a dynamical distinction between any two different
conformal frames (or unit systems) because they are generally
related via a spacetime dependent conversion factor. When the
universe goes through phase transition, one should incorporate
the resulting vacuum energy density to the model (\ref{2.1}) by
noting this dynamical distinction. We should therefore write the
action (\ref {2.1}) as
\begin{equation}
S=-\frac{1}{2} \int d^{4}x \sqrt{-g}~ \{g^{\mu\nu}
\nabla_{\mu}\phi \nabla_{\nu}\phi+\frac{1}{6}( R -2\bar{\Lambda}
e^{-2\sigma}) \phi^{2}\} \label{xy}\end{equation} where
$g^{\mu\nu}$ and $\phi$ are the metric tensor and the scalar
field in the cosmological frame.  These are related to
$\bar{g}^{\mu\nu}$ and $\bar{\phi}$ by the relations (\ref{2.6})
and (\ref{2.7}).  Here $\bar{\Lambda}$ is a typical mass scale
introduced by elementary particle physics measured in the quantum
frame. Note that since $\bar{\Lambda}$ carries the dimension of
squared mass it appears in the action (\ref{xy}) with an
exponential factor $e^{-2\sigma}$. The corresponding value of
$\bar{\Lambda}$ in the cosmological frame is $\Lambda$ which is
given by
\begin{equation}
\Lambda= \bar{\Lambda} e^{-2\sigma} \label{xz}\end{equation} As a
consequence various mass scales introduced by elementary particle
physics should have variable contributions to vacuum energy
density in the cosmological frame.\\ To study the evolution of
such mass scales we let the above action involve a kinetic term
for $\sigma$. In this way we consider $\sigma$ as a dynamical
field. This seems to be necessary to account for the dynamical
distinction between any two different unit systems. The action
(\ref{xy}) takes then the form
\begin{equation}
S=-\frac{1}{2} \int d^{4}x \sqrt{-g}~ \{g^{\mu\nu}
\nabla_{\mu}\phi \nabla_{\nu}\phi+(\frac{1}{6}( R-2 \bar{\Lambda}
e^{-2\sigma}) +\alpha g^{\mu\nu}\nabla_{\mu} \sigma
\nabla_{\nu}\sigma)\phi^2 \} \label{3.1.2}\end{equation} where
$\alpha$ is a dimensionless constant parameter.  This action can
now be used to describe a vacuum-dominated
universe since it excludes any matter contribution.  \\
Variation of (\ref{3.1.2}) with respect to $g^{\mu\nu}$, $\phi$
and $\sigma$ yields, respectively,
\begin{equation}
G_{\mu\nu}+\bar{\Lambda} e^{-2\sigma} g_{\mu\nu}=6\phi^{-2}
\tau_{\mu\nu} \label{3.1.3}\end{equation}
\begin{equation}
\Box \phi-\frac{1}{6}R \phi+\frac{1}{3}\bar{\Lambda} \phi
e^{-2\sigma}-\alpha \phi \nabla_{\gamma }\sigma \nabla^{\gamma
}\sigma =0 \label{3.1.4}\end{equation}
\begin{equation}
\frac{1}{\sqrt{-g}}\nabla_{\mu}(\sqrt{-g} \phi^{2} g^{\mu\nu}
\nabla_{\nu}\sigma) =\frac{1}{3\alpha} \bar{\Lambda} \phi^{2}
e^{-2\sigma} \label{3.1.5}\end{equation} where
\begin{equation}
\tau_{\mu\nu}=-(\nabla_{\mu }
\phi \nabla_{\nu}\phi-\frac{1}{2}g_{\mu \nu} \nabla_{\gamma  }\phi
\nabla^{\gamma  }\phi)+\frac{1}{6}(\nabla_{\mu }\nabla_{\nu}-g_{\mu\nu}\Box
)\phi^2-\alpha \phi^2 (\nabla_{\mu}\sigma \nabla_{\nu} \sigma -\frac{1}{2}
g_{\mu\nu} \nabla_{\gamma}\sigma \nabla^{\gamma}\sigma)
\end{equation}
The exponential coefficient of $\bar{\Lambda}$ emphasizes that
this mass scale belongs to a unit system which is different from
that used in cosmology. Intuitively, one expects that there should
be no distinction between the cosmological and the quantum unit
systems at sufficiently early times so that
\newpage
$$
e^{-2\sigma} \rightarrow 1~~~~~~~~~~~~~~~~~~~~~~~~~~
$$
\begin{equation}
\Lambda=\bar{\Lambda} e^{-2\sigma} \rightarrow \bar{\Lambda}~~~~~~
as~~~~~~ t\rightarrow 0 \label{c3.1.5}\end{equation} This can be
taken as an early-time boundary condition for the dynamical field
$\sigma$. In an expanding universe the distinction between these
two unit systems is expected to increase with time since all
cosmological scales enlarge as the universe expands.  Thus the
conformal factor $e^{2\sigma}$ must be an increasing function of
time. According to (\ref{xz}), this automatically provides us with
a dynamical mechanism for reducing the mass scale $\Lambda$ in
the cosmological frame.  It is important to note that this
mechanism
essentially works due to cosmic expansion. \\
Before studying this mechanism we would like to focus on the
behaviour of the field equations at early times.  To do this, we
apply the field equations to a homogeneous and isotropic universe.
In particular, we specialize to a spatially flat
Friedman-Robertson-Walker metric
\begin{equation}
ds^{2}=-dt^{2}+a^{2}(t)(dx^2+dy^2+dz^2)
\label{3.1.6}\end{equation} where $a(t)$ is the scale factor.
The homogeneity and isotropy require that the fields $\phi$ and
$\sigma$ be only functions of time.  The equations (\ref{3.1.3}),
(\ref{3.1.4}) and (\ref{3.1.5}) become
\begin{equation}
3(\frac{\dot{a}}{a})^2-\bar{\Lambda}
e^{-2\sigma}+3\frac{\dot{\phi}^2}{\phi^2}
+6\frac{\dot{a}}{a}\frac{\dot{\phi}}{\phi}+3\alpha
\dot{\sigma}^2=0 \label{3.1.7}\end{equation}
\begin{equation}
\frac{\ddot{\phi}}{\phi}+3\frac{\dot{a}}{a}\frac{\dot{\phi}}{\phi}
+(\frac{\ddot{a}}{a} +\frac{\dot{a^2}}{a^2})-\alpha
\dot{\sigma}^2-\frac{1}{3}\bar{\Lambda} e^{-2\sigma}=0
\label{3.1.8}\end{equation}
\begin{equation}
\ddot{\sigma}+(3\frac{\dot{a}}{a}+2\frac{\dot{\phi}}{\phi})\dot{\sigma}
+\frac{\bar{\Lambda}}{3\alpha}e^{-2\sigma} =0
\label{3.1.9}\end{equation} where the overdot indicates
differentiation with respect to $t$. We take\footnote{The
relations (\ref{3.1.11}) and (\ref{3.1.12}) indicate the
behaviour of $a(t)$ and $\phi(t)$ up to some proportionality
constants.  This is due to the fact that the field equations do
not change under scaling of $a(t)$ and $\phi(t)$.  These
proportionality constants, of course, do not have any role in our
arguments. }
\begin{equation}
a \sim e^{nt}
\label{3.1.11}\end{equation}
\begin{equation}
\phi \sim  e^{mt} \label{3.1.12}\end{equation}
and
\begin{equation}
\sigma=\xi t \label{x3.1.11}\end{equation} We substitute these and
the condition (\ref{c3.1.5}) into (\ref{3.1.7}), (\ref{3.1.8}) and
(\ref{3.1.9}) to obtain
\begin{equation}
n=\xi=\frac{1}{\sqrt{\alpha(4\alpha+1)}}\sqrt{\frac{1}{3}\Lambda}
\label{xx3.1.12}\end{equation}
\begin{equation}
m=(2\alpha-1)n
\end{equation}
These results indicate that the scale factor grows exponentially
and the spacetime geometry is described by the de Sitter metric.
This inflation continues to be a solution until $e^{-2\sigma}
\approx 1$ holds. After a sufficiently long time this
approximation is no longer valid and the inflation ends. The
point which we wish to make here is that our model does not
provide any contradiction with existence of an inflationary phase
at early times.  This is important since the basic idea of
inflation seems to be the only reasonable programme, suggested so
far, to resolve the cosmological
puzzles such as the flatness and the horizon problems \cite{guth}.\\
\subsection{Matter-dominated era}
To apply the model to a matter-dominated universe
we should first add a matter system in the action (\ref{3.1.2}).  We
write
\begin{equation}
S=-\frac{1}{2} \int d^{4}x \sqrt{-g}~ \{g^{\mu\nu}
\nabla_{\mu}\phi \nabla_{\nu}\phi+\phi^{2}(\frac{1}{6}(R-2
\bar{\Lambda} e^{-2\sigma}) +\alpha g^{\mu\nu}\nabla_{\mu} \sigma
\nabla_{\nu}\sigma)\}+S_{m}[g_{\mu\nu}]
\label{3.2.1}\end{equation} where $S_{m}[g_{\mu\nu}]$ is the
matter field action.  The gravitational equations for the action
(\ref{3.2.1}) will be
\begin{equation}
G_{\mu\nu}+\bar{\Lambda} e^{-2\sigma} g_{\mu\nu}=6\phi^{-2}
(T_{\mu\nu} +\tau_{\mu\nu}) \label{3.2.2}\end{equation} where
\begin{equation}
T_{\mu\nu}=\frac{2}{\sqrt{-g}}\frac{\delta}{\delta g^{\mu\nu}}
S_{m}[g_{\mu\nu}]
\end{equation}
The field equations of $\phi$ and $\sigma$ remain unchanged in the
presence of matter.  We may put
the equation (\ref{3.1.4}) into
the trace
of (\ref{3.2.2}) to obtain
\begin{equation}
T^{\gamma}_{\gamma}=\frac{1}{3}\bar{\Lambda} \phi^{2} e^{-2\sigma}
\label{3.2.3}\end{equation} which is a relation between the trace
of the matter stress-tensor and the vacuum energy density.  This
relation have important implications that we describe in the
following:\\ Let us first take $T_{\mu\nu}$ to be the
stress-tensor of a perfect fluid with energy density $\rho$ and
pressure $p$
\begin{equation}
T_{\mu\nu}=\rho u_{\mu} u_{\nu}+p(g_{\mu\nu}+u_{\mu}u_{\nu})
\label{3.2.3x}\end{equation} where $u_{\mu}$ is the four velocity
of the fluid.  For a pressureless fluid, (\ref{3.2.3}) takes the
form
\begin{equation}
\rho \phi^{-2} \sim \bar{\Lambda} e^{-2{\sigma}}
\label{3.2.5}\end{equation} In an expanding universe $\rho
\phi^{-2}$ decreases. The relation (\ref{3.2.5}) then predicts
that the same thing happens for $\Lambda$. Two important results
arise from this statement. Firstly, the expansion of the universe
induces the reduction of $\Lambda$. This requires that the
conformal factor $e^{2\sigma}$, or the $\sigma$ field, be an
increasing function of time.  In the previous subsection we
discussed that this is indeed the case in a vacuum-dominated
universe\footnote{Note that $\xi>0$ in the relation
(\ref{x3.1.11}).}. In this section we shall show that this is
also true
in a matter-dominated universe.\\
Secondly, the conformal invariance of our model does not allow to
incorporate naively a constant mass scale such as
$\bar{\Lambda}$.  Due to (\ref{3.2.3}), this would not be
dynamically consistent with the field equations. This emphasizes
the role of the dynamical distinction between the cosmological and
the quantum unit systems in our model.\\
We intend now to investigate the field equations in a
matter-dominated universe. For the metric (\ref{3.1.6}), the
equation (\ref{3.2.2}) becomes
\begin{equation}
3(\frac{\dot{a}}{a})^2-\bar{\Lambda} e^{-2\sigma}
+3\frac{\dot{\phi}^2}{\phi^2}
+6\frac{\dot{a}}{a}\frac{\dot{\phi}}{\phi}+3\alpha
\dot{\sigma}^2=6\phi^{-2} \rho \label{3.2.4}\end{equation} For
late times we take
\begin{equation}
a \sim (\frac{t}{t_{0}})^{v} \label{x3.2.5}\end{equation}
\begin{equation}
\phi=const. \label{3.2.6}\end{equation}
\begin{equation}
e^{\sigma}=\sigma_{0}t \label{3.2.7}\end{equation} where $t_{0}$
is the present age of the universe and $\sigma_{0}$ is a constant
with dimension of mass. Let us first estimate the gravitational
coupling. We substitute (\ref{3.2.6}) into the equation
(\ref{3.2.4}) to obtain
\begin{equation}
3H^2-\bar{\Lambda} e^{-2\sigma}+3\alpha
\dot{\sigma}^2=6\phi^{-2}\rho \label{3.2.8}\end{equation} where
$H=\frac{\dot{a}}{a}$ is the Hubble parameter. Using
(\ref{3.2.5}) this equation become
\begin{equation}
3H^2 +3\alpha \dot{\sigma}^2 \sim \phi^{-2}\rho \label{3.2.9}
\end{equation} From the relation (\ref{3.2.7}), one infers that at late times
$\dot{\sigma} \rightarrow t^{-1}\sim H$.  Thus (\ref{3.2.9})
reduces to
\begin{equation}
H^2 \sim \rho \phi^{-2} \label{3.2.10}\end{equation} Now we may
use the observational fact that \cite{17}
\begin{equation}
\rho \sim \rho_{c} \label{x3.2.12}\end{equation} with $\rho_{c}$
being the critical density
\begin{equation}
\rho_{c}=\frac{3H_{0}^{2}}{8\pi G} \label{xx3.2.12}\end{equation}
Here $G$ is the gravitational constant and $H_{0}$ is the Hubble
constant. From (\ref{3.2.10}) and (\ref{x3.2.12}), when $t
\rightarrow t_{0}$ we obtain
\begin{equation}
\phi^{-2} \sim G \label{3.2.17}\end{equation} Thus the constant
configuration of $\phi^{-2}$ at late times is given by the
gravitational constant. In this case the action (\ref{3.2.1})
reduces to
\begin{equation}
S=-\frac{1}{16\pi G}\int d^4 x \sqrt{-g}(R-2\bar{\Lambda}
e^{-2\sigma}+6\alpha g^{\mu\nu}\nabla_{\mu}\sigma
\nabla_{\nu}\sigma)+S_{m}[g_{\mu\nu}] \label{3.2.12}\end{equation}
This differs from the usual Einstein-Hilbert action in the sense
that it contains a dynamical field $\sigma$ and a varying
cosmological
term $\bar{\Lambda} e^{-2\sigma}$.\\
Variation of (\ref{3.2.12}) with respect to $g^{\mu\nu}$ and
$\sigma$ gives the field equations
\begin{equation}
G_{\mu\nu}+\bar{\Lambda}
e^{-2\sigma}g_{\mu\nu}+6\alpha(\nabla_{\mu}\sigma
\nabla_{\nu}\sigma -\frac{1}{2}g_{\mu\nu}\nabla_{\gamma}
\sigma\nabla^{\gamma}\sigma)=8\pi G T_{\mu\nu}
\label{3.2.13}\end{equation}
\begin{equation}
\Box\sigma=\frac{\bar{\Lambda}}{3\alpha}e^{-2\sigma}
\label{3.2.14}\end{equation} For the metric (\ref{3.1.6}) and the
matter stress-tensor (\ref{3.2.3x}), these equations become
\begin{equation} 3\frac{\dot{a}^2}{a^2}-\bar{\Lambda}
e^{-2\sigma}+3\alpha\dot{\sigma}=8\pi G\rho
\label{3.2.15}\end{equation}
\begin{equation}
\frac{\dot{a}^2}{a^2}+2\frac{\ddot{a}}{a}-\bar{\Lambda}
e^{-2\sigma} -3\alpha\dot{\sigma}=-8\pi G p
\label{3.2.16}\end{equation}
\begin{equation}
\ddot{\sigma}+3\frac{\dot{a}}{a}\dot{\sigma}+\frac{\bar{\Lambda}}{3\alpha}
e^{-2\sigma}=0 \label{3.2.17}\end{equation} The equation
(\ref{3.2.15}) together with (\ref{3.2.16}) gives
\begin{equation}
3\frac{\ddot{a}}{a}-\bar{\Lambda} e^{-2\sigma}-6\alpha
\dot{\sigma}^2=-4\pi G(\rho+3p) \label{3.2.18}\end{equation} For
$p=0$, we combine this with (\ref{3.2.15}) to obtain
\begin{equation}
\frac{\dot{a}^2}{a^2}+2\frac{\ddot{a}}{a}-3\alpha\dot{\sigma}^2-\bar{\Lambda}
e^{-2\sigma}=0 \label{3.2.19}\end{equation} Now if we substitute
(\ref{x3.2.5}) and (\ref{3.2.7}) in the equations (\ref{3.2.17})
and (\ref{3.2.19}) we obtain
\begin{equation}
v=\frac{2}{3}~~~~,~~~~-3\alpha
\label{3.2.20}\end{equation}
\begin{equation}
\sigma_{0}=\sqrt{\frac{\bar{\Lambda}}{3\alpha(1-3v)}}
\label{3.2.21}\end{equation} One can take the solution
$v=\frac{2}{3}$ for $\alpha<0$ since only in this case
$\sigma_{0}^2> 0$. This corresponds to the solution of the
standard cosmological model for the evolution of the scale factor
in a matter-dominated universe.\\ For $v=-3\alpha$, we obtain
accelerating solutions (with $v>1$) for $\alpha < -\frac{1}{3}$.
The deceleration parameter
\begin{equation}
q=-\frac{\ddot{a}a}{\dot{a}^2}
\end{equation}
is
\begin{equation}
q=\frac{1}{v}-1
\end{equation}
which is negative for $\alpha <-\frac{1}{3}$.\\  On the other
hand, in the cosmological frame we obtain
\begin{equation}
\Lambda=\bar{\Lambda} e^{-2\sigma} \sim t^{-2}
\label{3.2.22}\end{equation} where we have used (\ref{3.2.7}) and
(\ref{3.2.21}).  This result is consistent with the observational
bound.\\
We see that the conformal factor, or equivalently the $\sigma$
field, plays two important role in our model.  Firstly, evolution
of this dynamical field induced by cosmic expansion damps a large
effective cosmological constant.  Secondly, it plays the role of
a quintessence field that causes the universe to accelerate at
late times.
\section{Summary and discussion}
We have investigated the cosmological consequences of a conformal
invariant gravitational model which is assumed to hold during the
very early stages of evolution of the universe.  The conformal
invariance of the model allows us to formalize a theoretical
framework in which there exists a dynamical distinction between
the two unit systems used in cosmology and elementary particle
physics.  It is argued that when the universe goes through phase
transition the resulting large effective cosmological constant
$\bar{\Lambda}$ as a mass scale introduced by particle physics is
related to the corresponding mass scale in the cosmological frame
by $\Lambda =\bar{\Lambda} e^{-2\sigma}$. Thus all mass scales
introduced by particle physics should be considered as variable
in the cosmological frame \cite{bs}. This automatically suggests
a cancellation mechanism caused by expansion of the universe.  We
emphasize that this feature is also suggested by the relation
(\ref{3.2.3}) which is a dynamical consistency relation on the
trace of the matter stress-tensor in
our gravitational system. \\
The question which naturally arises is whether such a variable
cosmological term alters the standard picture of early history of
the universe.  To address this question, we have shown that there
exists a solution for the field equations at early times
exhibiting an exponential growth of the scale factor.  It is
important to note that in this model there is a natural exit of
the universe from this inflationary phase namely when
$e^{2\sigma} \approx 1$ does not hold due
to growth of $\sigma$.\\
Our primary interest is to explore the cosmological constant
problem.  We have shown that the asymptotic solution of the field
equations in the matter-dominated era
leads to the following consequences:\\
1) The relation (\ref{3.2.22}) indicates that the cosmological
constant in the cosmological frame $\Lambda$ is of the same order
of $t^{-2}$ which is consistent with the upper bound set by
observations. The smallness of the cosmological constant is
therefore related to the fact that the universe is old.\\
2) The gravitational coupling in the present state of the universe
is
given by $\phi^{-2} \sim G$.\\
3) The scale factor exhibits a late-time asymptotic power law
expansion $a\propto t^{v}$ with $v>1$.  This implies that the
universe is accelerating and $\sigma$ plays the role of a
quintessence field.  The acceleration of the universe is
generally achieved by negative values of $\alpha$ ($\alpha <
-\frac{1}{3}$).  This means that $\sigma$ is an ordinary massless
scalar field with positive energy density.

\newpage

\end{document}